\documentclass[twocolumn,showpacs,preprintnumbers,amsmath,nofootinbib,aps,superscriptaddress]{revtex4}

\usepackage{graphicx}
\usepackage{dcolumn}
\usepackage{bm}

\newcommand*{\no}{\noindent}
\newcommand*{\bea}{\begin{eqnarray}}
\newcommand*{\eea}{\end{eqnarray}}
\newcommand*{\be}{\begin{equation}}
\newcommand*{\ee}{\end{equation}}
\newcommand*{\pd}{\partial}

\newcommand*{\pref}[1]{(\ref{#1})}

\newcommand*{\nn}{\nonumber}

\newcommand*{\tr}{\mathrm{tr}}

\def\eq#1{(\ref{#1})}
\def\Eq#1{Eq.~(\ref{#1})}

\newcommand {\apgt} {\ {\raise-.5ex\hbox{$\buildrel>\over\sim$}}\ }
\newcommand {\aplt} {\ {\raise-.5ex\hbox{$\buildrel<\over\sim$}}\ }

\def\s0#1#2{\mbox{\small{$ \frac{#1}{#2} $}}}
\def\0#1#2{\frac{#1}{#2}}

\newcommand{\Tr}{\mathrm{Tr}}

\newcommand{\I}{\mathrm{i}}

\newcommand{\Nc}{N_{\text{c}}}

\begin{document}


\title{The gluon propagator close to criticality}

\author{Axel Maas}
\email{axelmaas@web.de}
\affiliation{Theoretical-Physical Institute, Friedrich-Schiller-University Jena, Max-Wien-Platz 1, D-07743 Jena, Germany}
\author{Jan M. Pawlowski}
\email{J.Pawlowski@thphys.uni-heidelberg.de}
\affiliation{Institute for Theoretical Physics, University of
  Heidelberg, Philosophenweg 16, D-69120 Heidelberg, Germany}
\affiliation{ExtreMe Matter Institute EMMI, GSI, Planckstr.~1, D-64291 Darmstadt, Germany}
\author{Lorenz von Smekal}
\email{lorenz.smekal@physik.tu-darmstadt.de}
\affiliation{Institut f\"ur Kernphysik, Technische Universit\"at Darmstadt, Schlossgartenstr.~2, D-64289 Darmstadt, Germany}
\author{Daniel Spielmann}
\email{D.Spielmann@ThPhys.Uni-Heidelberg.de}
\affiliation{Institute for Theoretical Physics, University of
  Heidelberg, Philosophenweg 16, D-69120 Heidelberg, Germany}
\affiliation{ExtreMe Matter Institute EMMI, GSI, Planckstr.~1, D-64291 Darmstadt, Germany}

\date{\today}

\begin{abstract}
  
  There are good reasons that the deconfinement phase transition of
  pure Yang-Mills theory at finite temperature should also be
  reflected in the behavior of gauge-fixed gluonic correlation
  functions. Understanding this in detail would provide another important
  example of how physical information can be extracted from gauge-dependent
  correlations, which is not always obvious. Therefore, herein we study
  the behavior of the Landau-gauge gluon propagator of pure $SU(2)$ across the
  phase transition in 2+1 and 3+1 dimensions in order to
  assess to what extend the corresponding critical behavior is
  reflected in these correlations. We discuss why it should emerge
  from a continuum perspective and test our expectations in lattice
  simulations. A comparison with $SU(3)$ furthermore reveals quite clear
  indications for a sensitivity of the gluon propagator
  to the order of the transition.  
	
\end{abstract}


\maketitle

\section{Introduction}

Our knowledge of even the most basic characteristic features of the QCD phase diagram is still rather uncertain. A rich structure of possible phases and transitions between them is being discussed in various regions, see, e.g.~\cite{Jacobs:2007dw,Andronic:2009gj,Leupold:2011zz}. Understanding how these phase transitions manifest themselves both theoretically and experimentally in a joint effort is the great challenge of strongly interacting matter research.

One of the well established properties of QCD with 2+1 flavors of dynamical quarks is a crossover at vanishing chemical potential in a temperature range of about 150-180 MeV, e.g.\
\cite{Bazavov:2010sb,Cheng:2009be,Borsanyi:2010bp,Pawlowski:2010ht}. However, many observables nevertheless appear to reflect the deconfinement phase transition that would occur if all quarks were sufficiently heavy. In fact, whether and where in the so-called Columbia plot the change from a genuine phase transition to a crossover occurs might still change when going beyond QCD as an isolated theory. In particular, if one also considers the quarks' fractional electric charges and couplings to more than one gauge group as in the Standard Model, a center-like symmetry reemerges which can break at the transition as in the quenched case \cite{vonSmekal:2010la}.

This finite-temperature deconfinement phase transition of the pure
$SU(N)$ Yang-Mills theory in $d+1$ dimensions follows the $Z_N$
symmetry breaking pattern of a $d$-dimensional $q$-state Potts model
with $q=N$. It is of second order for $N=2$ (Ising universality class)
in $d=2$ and $3$ and for $N=3$ in $d=2$. The self-duality of the
2-dimensional Potts models thereby manifests itself in the free
energies of the confining electric fluxes being mirror images around
criticality of those of center-vortex ensembles with twisted
boundary conditions
\cite{deForcrand:2001nd,vonSmekal:2010xz,Strodthoff:2010dz}.

Another tool to determine the dynamics of order parameters for
the Yang-Mills phase transition, in particular the Polyakov loop,  
the chiral condensate 
and to some extent thermodynamic bulk
properties, have been gauge-fixed correlation functions, in particular in the Landau gauge. While it has been known since decades how the chiral phase transition and in particular its critical properties manifest themselves directly in the
gauge-fixed quark correlation functions, it has only recently become clear how to compute the non-perturbative effective potential of deconfinement
order parameters from the gauge-fixed propagators 
\cite{Braun:2007bx,Marhauser:2008fz,Braun:2010cy,Braun:2009gm,Fischer:2009wc,Fischer:2009gk,Fischer:2010fx,Pawlowski:2010ht}. 
These analyses entail that already the gauge-fixed propagators carry the critical 
properties of the confinement-deconfinement phase transition, which has been used explicitly in 
\cite{Marhauser:2008fz}. These findings have also been studied using lattice simulations, 
see in particular \cite{Cucchieri:2007ta,Fischer:2010fx}, though systematic errors are
still a matter of ongoing research
\cite{Cucchieri:2007ta,Fischer:2010fx,Bornyakov:2010nc,Cucchieri:2011di,Aouane:2011fv}.

In the present paper we extend these investigations. We first
review in Section \ref{sec:crit} how critical
properties of the phase transition are incorporated in the gauge-fixed
correlation functions. We argue that the gluon propagator should be
directly sensitive to the phase transition and, in particular, that it should
reflect critical scaling. In Section~\ref{sec:lat} the behavior expected from 
these arguments is compared to results from lattice simulations in
three and four dimensions for $SU(2)$ Yang-Mills theory. In both cases the phase
transitions are of second order, but one should be able to distinguish
the different critical exponents corresponding the $2d$ and $3d$ Ising
universality classes, respectively. We also briefly revisit the
first-order case of four-dimensional $SU(3)$ Yang-Mills theory to assess
the sensitivity of the gluon propagator to the order of the transition
in Section \ref{sec:su3}. Our summary and conclusion are given in 
Section \ref{sec:sum}. Some additional results on the momentum
dependence of the propagators are deferred to Appendix
\ref{app:full}, while an extended discussion of the systematic and statistical errors
is given in Appendices \ref{app:sys} and \ref{app:stat}, respectively.

\section{Critical behavior and the gluon propagator}\label{sec:crit}

The second-order finite-temperature deconfinement phase transitions of
$SU(2)$ Yang-Mills theory in four, and both $SU(2)$ and $SU(3)$ in three
dimensions, are characterized by correlation lengths which diverge
at the critical temperature in the infinite volume limit. As a 
consequence, only the long-range properties of the theory matter near
criticality. These are determined by symmetries and dimensionality of
the system, and the concepts of universality and scaling apply. 
The second-order $SU(N)$ cases belong to Potts/Ising universality
classes in two and three dimensions as mentioned above. 
In particular, the $SU(2)$ Yang-Mills theories in three and four
dimensions that we are mainly interested in here, belong to the classes of
the two-dimensional and three-dimensional Ising models, respectively.

Critical behavior manifests itself in singularities of thermodynamic
functions at the critical temperature $T_c$ (and vanishing external field). 
In particular, these singularities show as characteristic non-integer powers
of the reduced temperature
\be
t=\left(\frac{T}{T_c}-1\right)\label{crit}.
\ee
\no 
The critical exponents in these power laws only depend on
symmetry-breaking pattern and dimension, and are thus a characteristic
feature of a universality class. Scaling and hyper-scaling (in four
dimensions or less) entail that all critical exponents can be
expressed in terms of two independent ones, say $\nu $ and $\eta$.
The first is the critical exponent of the diverging correlation length
$\xi$, i.e., 
\begin{eqnarray}\label{eq:correlationslength}
\xi \propto |t|^{-\nu}\,, 
\end{eqnarray}   
close to a second-order phase transition. The second is defined
from the (in general renormalization-group-dependent) connected propagator $G(p)$ of the order parameter, whose
zero momentum contribution defines the susceptibility which diverges with
exponent $\gamma$, 
\begin{eqnarray} \label{eq:prop0} 
G(0) \propto |t|^{-\gamma}\,,\quad {\rm with}\quad \gamma =\nu(2 -\eta)\,. 
\end{eqnarray} 
In the broken phase, an order parameter $M$ scales with the exponent $\beta$,
\be\label{eq:orderparameter}
M(t)\propto |t|^\beta
\quad {\rm with}\quad \beta =\012 \nu(d-2 +\eta)\,.
\ee 
Usually $\beta$ is bound by its mean field value, i.e.
$\beta\le 1/2$, and the derivative of $M(t)$ with respect to $t$
therefore diverges at the critical temperature. 

A natural order parameter for confinement is provided by the Polyakov loop, 
\begin{equation}\label{eq:Polloop}
  L(\vec x)=\frac{1}{\Nc} \tr\, \text{P}\, \exp\left( { \I g\int_0^{1/T} 
    d \tau\, {A}_0(\tau,\vec x)}\right)\,. 
\end{equation}
It depends on the zero component $A_0$ of the gauge field, and its
correlation length is inversely related to the string tension, 
i.e., $\xi_- = T/\sigma(T) $ for $T < T_c$. Therefore, with
second-order transition, $\sigma \propto (-t)^\nu$ for $t \to
0^-$.\footnote{Above $T_c$, the Polyakov-loop correlation length
  $\xi_+$ is related to the dual string or interface tension
  $\tilde\sigma \propto (\xi_+)^{1-d} $, a well-defined order
  parameter for center-symmetry breaking
  \cite{DeForcrand:2001dp}. With second-order transition,
  $\tilde\sigma \propto t^\mu $ for $ t \to 0^+$, and (hyper-scaling)
  $\mu = (d-1)\nu $.}  

In a gauge-fixed approach the theory is described in terms of, in general, 
gauge-dependent correlation functions, i.e.\ $N$-point functions of
the gauge field. These correlation functions have to contain the
information about the physics at criticality, even though they change
under gauge transformations. 

Indeed it has been shown how to extract the expectation value of the
(logarithm of the) Polyakov loop from the gauge-fixed propagators in
general gauges in terms of the full effective potential $V_{\rm eff}$
of $\langle A_0\rangle$
\cite{Braun:2007bx,Marhauser:2008fz,Braun:2010cy}.  The value of
$\langle A_0\rangle$ at the minimum of $V_{\rm eff}$ directly relates
to the expectation values of the eigenvalues of the (logarithm of the)
Polyakov loop, it is gauge invariant and also provides an order
parameter. Other recent examples are dual condensates 
\cite{Gattringer:2006ci,Synatschke:2007bz,Bilgici:2008qy,Zhang:2010ui,Bilgici:2009tx}, 
which can likewise be computed from gauge-fixed correlation functions 
\cite{Fischer:2009wc,Fischer:2009gk,Braun:2009gm,Fischer:2010fx,Fischer:2011mz}, and in the quenched case determine the Polyakov loop directly.

In \cite{Braun:2007bx,Braun:2010cy} the potential has been computed in
Landau gauge; the critical temperature agrees well with lattice
results. In \cite{Marhauser:2008fz} the same result was obtained within a computation in Polyakov gauge, a necessary requirement for the results to be gauge-independent as they must be.
Moreover it has been
verified that in $SU(2)$ the potential and hence the order parameter
does indeed show the scaling of the Ising universality class.

An interesting consequence of these computations and the related
formal analysis is that the gauge-fixed propagators necessarily also show
critical scaling. Here we briefly review the corresponding 
results in \cite{Braun:2007bx,Marhauser:2008fz,Braun:2010cy}.  In
\cite{Braun:2007bx} it has been shown that the effective potential of
$\langle A_0\rangle$ in a general gauge can be computed solely from the
ghost and gluon propagators in the constant background $\bar A_0$:
\begin{eqnarray}\nonumber 
  \partial_k V_{\rm eff,k}(\bar A_0) &= &\0{1}{2} 
  \Tr\,\langle A^a_\mu A^b_\nu
  \rangle^{\ }_{\bar A_0} \partial_k R_{k,\nu\mu}^{ba} \\
  &&  - \0{1}{2} \Tr\, \langle \bar C^a 
  C^b\rangle^{\ }_{\bar A_0}  \partial_k R_k^{ba}\,, 
\label{eq:flow} \end{eqnarray}
where $k$ is an infrared cut-off scale implemented by the regulator $R_k^{ab}$ in the functional renormalization group equations. At large cut-off scales the
effective potential vanishes. Hence, integrating the above equation
\eq{eq:flow} from the trivial potential at $k\to \infty$ to $k=0$
provides us with $V_{\rm eff}(\bar A_0) = V_{\rm eff,k=0}(\bar A_0) $ the
full non-perturbative effective potential of $\langle A_0\rangle$.

Since $V_{\rm eff}(\bar A_0)$ is the potential of the order parameter
$\langle \log L\rangle $, it has to show critical scaling (if the
transition is of second order). Consequently, also the propagators
will be sensitive to critical scaling, in general.

The above setting even allows us to extract the scaling of the
longitudinal gluon. Its propagator schematically is of the form,  
\begin{eqnarray}\label{eq:funproplong} 
  \langle A_L A_L  \rangle^{\ }_{\bar A_0} =\Pi_{L} 
\frac{1}{(-\bar D_0^2+\vec p^2 )Z_{L}}\Pi_{L} \,,
\end{eqnarray}  
where $\Pi_{L}$ stands for the projection operator longitudinal to the
heatbath in the presence of the background $\bar A_0$, and $
i\bar D_0=p_0+ g \bar A_0$. The prefactor $Z_{L}=Z_{L}(-\bar
D_0^2,\vec p^2;\bar A_0)$ depends on $\bar A_0$, $\bar D_0^2$ and $\vec p^2$ separately and
reduces to the well-known longitudinal (inverse) Landau gauge dressing function for
$\bar A_0=0$. 

The solution $a_0$ of the equations of motion for $\bar A_0$ is the order parameter related to 
$\langle \log L \rangle$ and derives from
\begin{eqnarray}\label{eq:EoMA0}
\left. \frac{\partial V_{\rm eff}}{\partial \bar A_0} \right|_{\bar A_0=a_0}=0 \,.
\end{eqnarray}
Hence it scales according to \eq{eq:orderparameter}, and the effective
potential also shows critical scaling, i.e. 
\begin{eqnarray}\label{eq:nufromV}
\frac{\partial^2 V_{\rm eff}}{\partial \bar A_0^2} (a_0)\propto |t|^{\gamma}\,. 
\end{eqnarray}
with $\gamma=\nu(2-\eta)$, see \eq{eq:prop0}. Up to possible 
dependencies on the background field, the
inverse longitudinal gluon propagator is proportional to
\eq{eq:nufromV}, $\langle A_L A_L \rangle^{-1} \propto \partial^2
V_{\rm eff}/\partial \bar A_0^2$.  In other words, the mass gap is
given by the curvature of $V_{\rm eff}(\bar A_0)$ in
\eq{eq:nufromV}, and hence determines the singular behavior of the
gluon propagator at vanishing frequency and momentum, i.e.
\begin{eqnarray} \label{eq:singscaling}
%
  \langle A_L A_L  \rangle^{\ }_{\bar A_0=a_0}(p=0)  \propto
  |t|^{-\gamma}\,.  
\end{eqnarray} 
We conclude that at least the longitudinal or electric propagator
\eq{eq:funproplong} scales as the propagator of an order parameter,
see \eq{eq:prop0}. In \cite{Marhauser:2008fz} this has in fact been used to
determine the critical scaling directly from the longitudinal gluon propagator. 

Note however, that a similar analysis cannot be made for
transverse gluons and ghosts. This is due to the fact that their
inverse propagators contain no $\bar{A}_0$-derivatives and hence no dependence
on the curvature term in \eq{eq:nufromV}. 
Consequently, we cannot infer from this simple argument whether one should
see critical scaling in the transverse gluon and ghost
propagators or not. However, it is suggestive that it at best arises in 
sub-leading order, that is, that singularities might appear in higher
$T$-derivatives. This is supported by lattice results, see \cite{Cucchieri:2007ta,Maas:2011se,Fischer:2010fx,Bornyakov:2010nc} and appendix \ref{app:stat}.

Now we turn to lattice Landau gauge and, in particular, discuss the
differences of simulations there to the background Landau gauge
scenario used above. The screening mass of a two-point correlation
function, the propagator $D(p)$, defined as 
\be
M_s=\frac{1}{\sqrt{D(0)}}\label{eq:screening}\,, 
\ee 
is a candidate for critical
behavior. Note that the corresponding pole mass
\cite{Maas:2011se}, if it exists, will in general only be sensitive to
correlation lengths up to its inverse size. Since all results
available so far \cite{Maas:2011se} seem to exclude a zero pole mass
of the gluons, it is therefore to be expected that the pole mass will
not exhibit critical behavior.

The temperature-dependence of the electric
and magnetic gluon propagators and the ghost propagator in lattice
Landau gauge has been investigated previously, for details see
\cite{Cucchieri:2007ta,Fischer:2010fx,Bornyakov:2010nc,Cucchieri:2011di,Maas:2005hs,Maas:2004se,Gruter:2004bb,Zahed:1999tg,Cucchieri:2000cy,Cucchieri:2001tw,Karsch:1994xh,Chernodub:2007rn,Aouane:2011fv,Fister:2011ym}. One
observes that only the electric propagator is sensitive to the phase
transition \cite{Cucchieri:2007ta,Fischer:2010fx,Aouane:2011fv}. However, the
available lattice data do not yet allow to draw firm conclusions
about a possible scaling behavior at criticality. Nevertheless, the 
susceptibility $\chi^E$ defined as the temperature derivative of 
its screening mass $M_s^E$,
\be
\chi^E=\frac{\pd M_s^E}{\pd T}\; ,\nn 
\ee \no 
is sufficiently sensitive to determine the phase transition
temperature within the systematic errors, see \cite{Fischer:2010fx}  and 
Appendix~\ref{app:sys}.

The aim of the present investigation is the further assessment of the critical
behavior. This will be performed using lattice simulations in three
and four dimensions for the gauge group SU(2). In four dimensions, for this purpose, a significantly refined temperature mesh and increased statistics were used compared to the previous study \cite{Fischer:2010fx}. The three-dimensional
system, investigated here for the first time, has the advantage, compared to the four-dimensional one, that
the expected critical exponents are larger, and therefore the signal
might be stronger. 

\begin{figure}
\includegraphics[width=\linewidth]{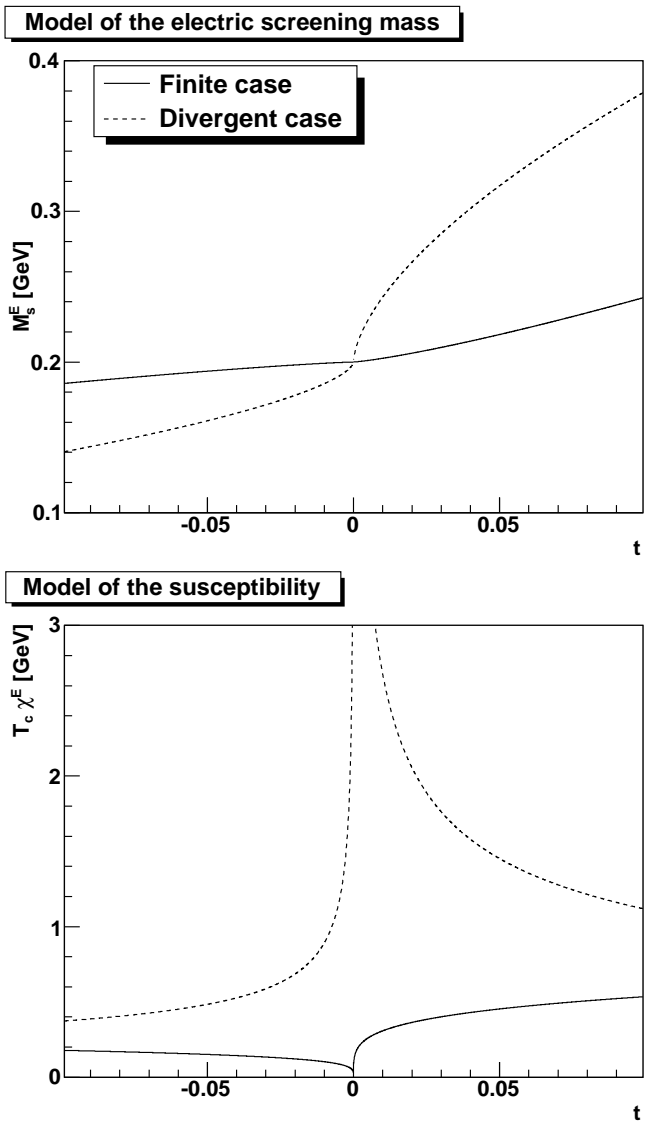}
\caption{\label{fig:cartoon}Cartoon of the electric screening mass and
  susceptibilities of the form \pref{eq:fullmass} and
  \pref{eq:fullmassopt}, with critical exponents $\gamma/2$ (divergent
  case) and $\gamma$ (finite case) with $\gamma=1.2372(5)$, i.\ e.,
  with the 3-dimensional Ising universality class. The non-universal
  constants are $m_{\rm gribov}=0.2$ GeV, $a=-1/4$ GeV and $b=3/4$
  GeV.}
\end{figure}

In lattice Landau gauge, naively, one would expect a similar behavior as
for the background Landau gauge discussed above. However, the two
differ in at least one important qualitative aspect. All currently employed
practical implementations of the lattice Landau gauge, including the
one used herein, exhibit a screening mass. Given its non-perturbative origin and appearance in a gauge-fixed correlator, it will be referred to as a Gribov mass \cite{Gribov:1977wm}\footnote{Note that in the original work \cite{Gribov:1977wm} the so defined screening mass was infinite.}, $m_{\rm gribov}$. This mass is
present at all temperatures and hence interferes with critical
scaling.  This entails that the inverse propagator can still have
a mass-gap at criticality. It is important to emphasize that the
effective potential $V_{\rm eff}$ for the order parameter $a_0$ in
such a gauge still satisfies \eq{eq:flow}. Hence, the propagators must
reflect critical scaling nevertheless. This implies that 
the screening mass $M^E_s$  of the electric propagator in
\eq{eq:screening} contains both, a regular Gribov mass contribution as
well as a singular critical one. From \Eq{eq:flow} one concludes that
the critical contribution is determined by the exponent $\gamma$. 
Unfortunately, this general argument is not sufficient to disentangle
the two contributions in a unique way. The two perhaps most natural 
candidates to explore at least as approximate descriptions near
criticality might be to either simply add the two mass contributions
directly, i.e.  
\begin{eqnarray} \label{eq:fullmass}
M_s^E(t)= m_{\rm gribov}+ a_{\pm}\, |t|^{\gamma/2}\, ,
\end{eqnarray} 
where $a_+=a$ is a non-universal coefficient for temperature $t>0$ and
$a_-=b$ for temperatures $t<0$; or, alternatively, add two
corresponding self-energy contributions to the inverse propagator
which would lead to a sum of squares for the total mass, i.e.
\begin{eqnarray} \label{eq:fullmassopt}
M_s^E(t)= \sqrt{m_{\rm gribov}^2+ a^2_{\pm}\, |t|^{\gamma}}\,.
\end{eqnarray} 
A simply testable criterion to at least discriminate these two cases is
provided by the corresponding susceptibilities  $\chi^E$ which near
criticality in the first case \eq{eq:fullmass} behaves as  $\chi^E
\propto |t|^{\gamma/2-1}$, while with \eq{eq:fullmassopt}, $\chi^E
\propto |t|^{\gamma-1}$. For $1 < \gamma < 2$, in the $2d$ and $3d$
Ising universality classes one has $\gamma = 7/4=1.75$ and $\gamma =
1.2372(5)$, respectively \cite{Pelissetto:2000ek}. This implies that 
\eq{eq:fullmass} leads to a susceptibility 
 $\chi^E= \partial_t M_s^E$ which would diverge  
 at the critical temperature (in the infinite volume limit), while
 \eq{eq:fullmassopt} would lead to a vanishing one. This is
 illustrated by a cartoon of the four-dimensional case in Figure
 \ref{fig:cartoon}. 

While there will of course neither be a divergence nor a strict zero
in  $\chi^E$ in a finite volume, the lattice results clearly favor
an enhancement of  $\chi^E$ near $T_c$ corresponding to \eq{eq:fullmass}
rather than a suppression, as can be seen from comparing the cartoon
in Figure \ref{fig:cartoon} with Figures \ref{fig:3d-em} and
\ref{fig:4d-em} below.  So we conclude that  \eq{eq:fullmassopt}
can relatively safely be ruled out. A more conclusive study of the
expected behavior based on a self-consistency analysis of \eq{eq:flow}
would be clearly desirable but will be postponed to future work.  

Furthermore, because the Gribov mass itself is significantly
volume dependent
\cite{Maas:2011se,Sternbeck:2007ug,Bogolubsky:2009dc,Cucchieri:2007rg,Cucchieri:2008fc},
it is not possible at present to clearly identify the critical
behavior in the electric screening mass or its temperature derivative
from a finite-size-scaling analysis of lattice results.
The Gribov mass creates a mass gap whose volume dependence especially
at criticality is yet to be determined. A significant amount of
different volumes, and possibly discretizations, will be necessary to 
disentangle the two intertwined finite-volume effects on the electric
screening mass, from critical scaling and from the Gribov mass, requiring
substantially more resources than presently available to us.

Finally, it should be noted that the ans\"atze \pref{eq:fullmass} and \pref{eq:fullmassopt} are expected to be only valid in the critical region. The Gribov mass in these fits will therefore not coincide with the one at zero temperature, as further, sub-leading temperature-dependent contributions are neglected.

\section{Lattice results}\label{sec:lat}

The gluon propagator is determined using standard lattice methods, 
for details see \cite{Cucchieri:2006tf,Cucchieri:2007ta}. We use the minimal
Landau gauge \cite{Maas:2011se} with the  implications
discussed in Section \ref{sec:crit}. In particular, the gluon
propagator exhibits a finite Gribov screening mass already at zero temperature,
while the ghost behaves essentially tree-level-like
\cite{Sternbeck:2007ug,Cucchieri:2007md,Bogolubsky:2007ud,Bogolubsky:2009dc,Cucchieri:2007rg,Cucchieri:2008fc}. In
four dimensions, in addition to the data from \cite{Fischer:2010fx},
further results using a finer discretization at the price of smaller
physical volumes, are used. In three dimensions, two different
discretizations are used as well. For the main purpose of this work
only the electric screening mass is of importance. For completeness,
also the finite momentum results are reported, but relegated to 
Appendix \ref{app:full}.

For setting the scale a string tension of $(440$ MeV$)^2$ has been used, based on the $a(\beta)$ values determined in \cite{Teper:1998te} for three dimensions and in \cite{Fingberg:1992ju} for four dimensions. This particular value was chosen to ensure comparability to the previous investigations \cite{Cucchieri:2007ta,Fischer:2010fx}. The critical $\beta$ values are taken from \cite{Edwards:2009qw} for three dimensions and from \cite{Fingberg:1992ju} for four dimensions. A discussion of systematic effects due to the scale setting can be found in Appendix \ref{app:sys:scale}.

\subsection{Three dimensions}

\begin{figure}
\includegraphics[width=\linewidth]{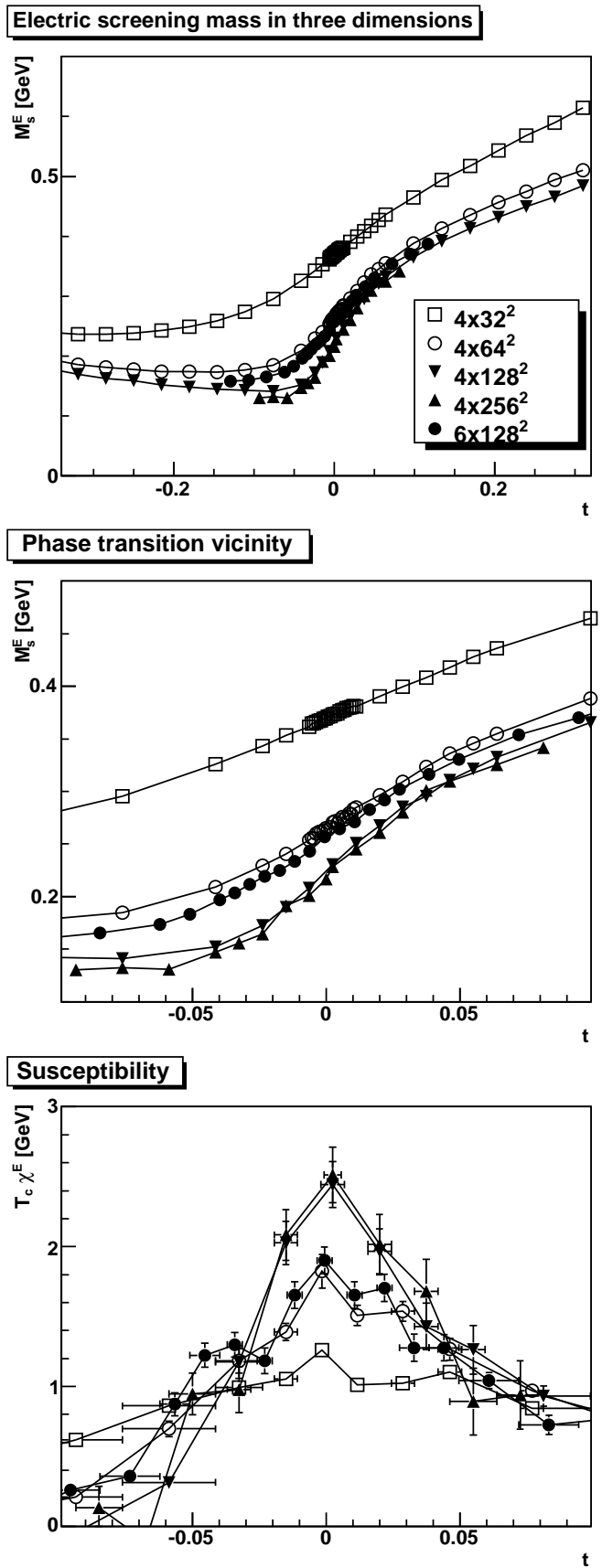}
\caption{\label{fig:3d-em}The electric screening mass (top panel and zoomed in in the phase transition region in the middle panel) as a function of the reduced temperature for various lattice sizes. The bottom panel shows the corresponding susceptibility, where some temperature points have been dropped to reduce the statistical errors.}
\end{figure}

The results for the electric screening mass, as well as its
susceptibility in three dimensions are shown in Figure
\ref{fig:3d-em}. It is clearly visible that the maximum of the
susceptibility marks the critical temperature, which has been
determined independently using analyses of the free energy and quantities derived from it \cite{Fingberg:1992ju,Edwards:2009qw}. Furthermore, a sizable volume scaling is
seen for the smaller spatial volumes with $N_t=4$ time
slices. For the largest spatial extent the peak in the
susceptibility only becomes somewhat sharper, but not significantly
higher. This might be  due to the mixing with the Gribov
mass as discussed in Section \ref{sec:crit}. Note, however, that the singular
contribution to  $\chi^E= \partial_t M_s^E \propto |t|^{\gamma/2-1} $ 
from our model \eq{eq:fullmass} would predict a finite-size $L$ scaling
of the maximum as $\chi^E_\mathrm{max} \propto L^{(1-\gamma/2)/\nu} =
L^{1/8}$ for three-dimensional $SU(2)$, corresponding to a factor $2^{1/8} $ or
about only $9\%$ increase with doubling the spatial lattice size which
is still within errors. Furthermore, increasing $N_t$ from 4 to 6,
which effectively reduces the physical $L$ for equal spatial lattice
sizes by $2/3$, we observe that the results remain qualitatively
unchanged, at least for the two values checked here. The details might
still be influenced by lattice artifacts. 

We reiterate that the screening mass, which is (semi)positive by definition,  
is non-zero at the phase transition. Moreover, because the peak in its
temperature derivative $\chi^E$ marks the phase transition, the
screening mass does not have a minimum there, but a point of
inflection.\footnote{From the coarser temperature resolution in
  \cite{Fischer:2010fx} it was suggested that the transition
  occurred at the maximum of the electric gluon propagator at zero
  momentum, i.e. for minimal $M_s^E$, which is slightly off.   
  However, this merely implies that the phase transition temperature should be
  set to the peak in the susceptibility $\chi^E$ instead. Otherwise the
  results of \cite{Fischer:2010fx} remain unchanged.} 
It is also well established that there is a finite screening mass in
minimal Landau gauge at zero temperature \cite{Cucchieri:2007rg}. 
The possibility that it remains constant throughout all
temperatures up to the phase transition, and that it starts to
increase abruptly there, would imply that the non-universal
coefficient $a_- = b=0$ and is included here. For our fits we thus
assume a constant contribution corresponding to a Gribov mass 
introduced by the gauge-fixing, which should be
temperature-independent, to which we add a form resembling a critical
behavior as in \pref{eq:fullmass}. Subleading contributions around
criticality will be neglected, leading to the Ansatz for the screening
mass, 
\be
M_s^E(t)=m_{\rm gribov}+\theta(t) \,a \, t^\frac{\gamma_+}{2}+\theta(-t)\, b\,
(-t)^\frac{\gamma_-}{2}\label{fit} ,
\ee
\no where we have furthermore introduced independent exponents
$\gamma_+$ and $\gamma_-$ above and  below
$T_c$ in order to assess to what extend the results are consistent with a
unique $\gamma_+ = \gamma_-$. As a side remark note that the
Ansatz \eq{fit} for small $t$ in principle also includes the form
in \pref{eq:fullmassopt} for which one should obtain $\gamma_+ =
\gamma_- = 2\gamma$. 

\begin{table*}
 \caption{\label{tab:3d}The fit parameters for the fit \pref{fit} for the three-dimensional case. For the determination and placement of the statistical errors see Appendix \ref{app:stat}. Only the change in the last digits is indicated.  See Appendix \ref{app:sys} for a discussion of systematic errors.}
 \begin{tabular}{|c|c|c|c|c|c|}
 \hline
 Lattice    & $m_{\rm gribov}$ [GeV] & $a$ [GeV] & $b$ [GeV] & $\gamma_+$ & $\gamma_-$ \cr
 \hline
 $4\times32^{2}$  & 0.37019$_{+3}^{-2}$ & 0.85$_{+5}^{-4}$ & -0.78$_{-4}^{+4}$ & 1.88$_{+4}^{-4}$ & 1.81$_{+4}^{-4}$ \cr
 \hline
 $4\times64^{2}$  & 0.2637$_{+6}^{-6}$ & 0.90$_{+5}^{-4}$ & -0.67$_{-3}^{+3}$ & 1.68$_{+4}^{-4}$ & 1.63$_{+2}^{-2}$ \cr
 \hline
 $4\times128^{2}$ & 0.221$_{+3}^{-3}$ & 0.735$_{+7}^{-6}$ & -0.368$_{-14}^{+13}$ & 1.39$_{+10}^{-9}$ & 1.134$_{+2}^{+3}$ \cr
 \hline
 $4\times256^{2}$ & 0.218$_{+3}^{-3}$ & 0.73$_{+11}^{-8}$ & -0.33$_{-3}^{+2}$ & 1.39$_{+12}^{-10}$ & 1.02$_{+3}^{-3}$ \cr
 \hline
 $6\times128^{2}$ & 0.2571$_{-5}^{+5}$ & 0.80$_{+2}^{-2}$ & -0.54$_{-3}^{+3}$ & 1.63$_{+16}^{-15}$ & 1.38$_{+4}^{-4}$ \cr
 \hline
 \end{tabular}
\end{table*}

\begin{figure}
\includegraphics[width=\linewidth]{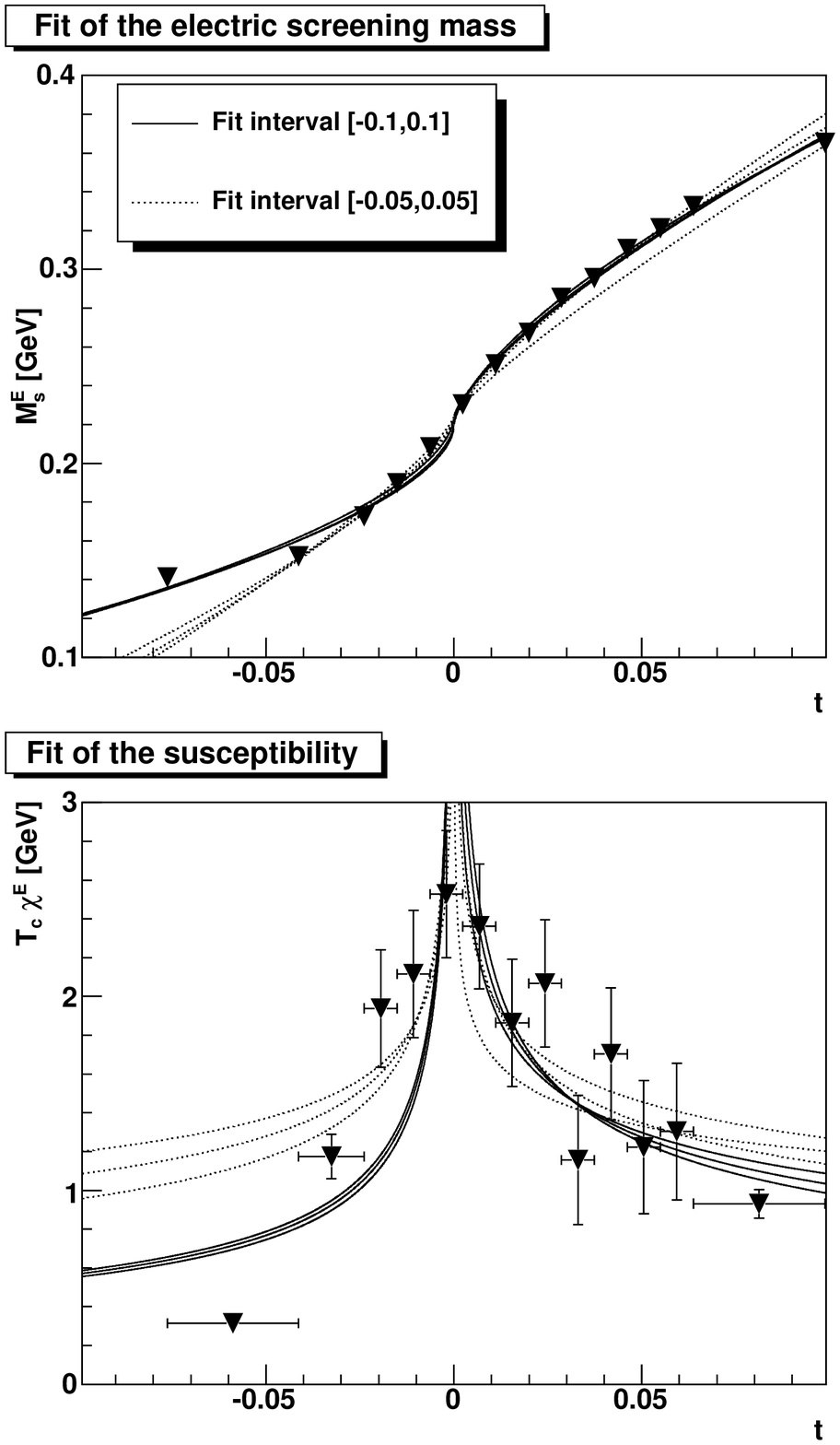}
\caption{\label{fig:3d-fit}The quality of the fits compared to the data from the $4\times128^2$ lattice. The standard fit uses the values from Table \ref{tab:3d}. As an alternative, a fit using only a more restricted reduced temperature interval $[-0.05,0.05]$ is also shown (dotted line), which yields $\gamma_-=1.52_{+9}^{-10}$ and $\gamma_+=1.61_{+14}^{-12}$. Fits have been performed on the screening mass using all data points, and the derivatives have then been calculated analytically. The error bands are also shown.}
\end{figure}

Performing the fits in the reduced-temperature interval $[-0.1,0.1]$ yields the results shown in table \ref{tab:3d}. An example of the quality of the fit is shown in Figure \ref{fig:3d-fit}. There are a number of observations:

First, the fit form \pref{fit} describes the screening mass rather
accurately, and within the statistical errors also the susceptibility in a
satisfactory way. The resulting exponents are all in the range
$1<\gamma<2$ (and thus inconsistent with \pref{eq:fullmassopt}), but they
do not agree with one another very well, which is only slightly improved if the
fit interval is further reduced, assuming a smaller
critical region.   

Given that the fits provide a rather good description of both the screening mass and the susceptibility there are a number of possible reasons for the observed non-singularity in the data itself. The first is that the volumes are just too small and the discretization too coarse to be close enough to a critical behavior. Given that the critical behavior is well visible on comparable lattices for other observables \cite{Edwards:2009qw}, this appears at first glance to be a rather unlikely explanation. However, it is possible that the electric gluon propagator lags significantly in the development of the critical behavior behind those other observables. Indeed, the gluon propagator requires already at zero temperature very large volumes to reach even its qualitative infinite-volume behavior \cite{Sternbeck:2007ug,Bogolubsky:2009dc,Cucchieri:2007rg,Cucchieri:2008fc}. This appears to be true also at finite temperature \cite{Cucchieri:2007ta,Cucchieri:2011di}. Therefore this appears to be a possible explanation.

The second possibility is that the critical region is very narrow. Then it would be necessary to investigate the region around the critical temperature with a much finer mesh. In this case also significantly enhanced statistics would be necessary, since the numerical derivative for the determination of the susceptibility requires that the statistical error must be much smaller than the distance between two points.

Finally, it is of course still possible that the electric gluon
propagator simply does not show critical behavior. At the moment this
cannot entirely be excluded, despite the general continuum arguments
in favor given in Section \ref{sec:crit}. This could be due, e.g.\ 
to a more complicated interference with the Gribov mass than the two
possibilities considered in  \pref{eq:fullmass} and  \pref{eq:fullmassopt}.  
 An alternative possibility will be discussed in Section
 \ref{sec:alt} below. However, since the behavior of the Polyakov loop can
 be derived from the propagators \cite{Fischer:2010fx}, this appears
 somewhat unlikely, though by far not impossible. 

\subsection{Four dimensions}

\begin{figure}
\includegraphics[width=\linewidth]{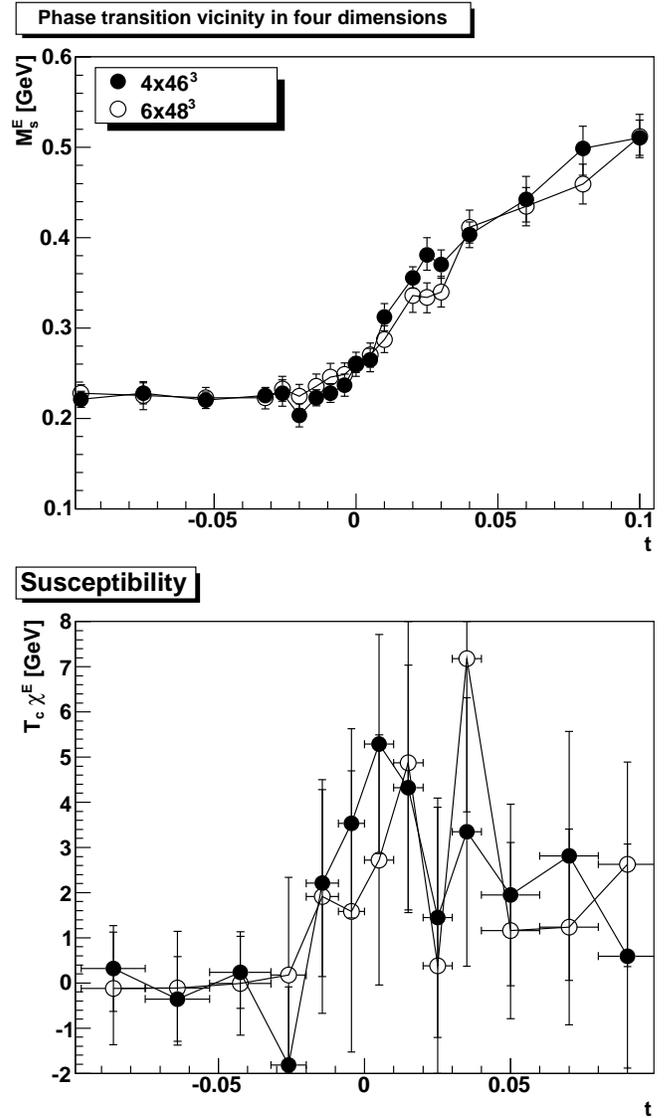}
\caption{\label{fig:4d-em}The electric screening mass (top panel) as a function of the reduced temperature for two different lattice sizes. The bottom panel shows the corresponding susceptibility, where some temperature points have been dropped to reduce the statistical errors.}
\end{figure}

Essentially the same qualitative behavior, though within a different
universality class, is expected in four dimensions. The corresponding
electric screening mass and its susceptibility are shown in Figure
\ref{fig:4d-em}. The qualitative picture is the same as in three
dimensions. Note, however, that because the critical temperature is
about one third lower in four as compared to three dimensions, the
physical volumes are about 50\% larger here than they would be for the
same number of lattice points in three dimensions. Even though our
volumes are nevertheless still smaller than in the three-dimensional
case, the susceptibility is significantly more peaked, showing that
the phase transition leaves a much more significant imprint. However,
the peak of the susceptibility is somewhat displaced as compared to the
phase transition point. One possible reason could be systematic errors in
the scale setting. Other, equally possible, reasons may be statistical fluctuations, finite volume effects or discretization effects. Permitting even only a systematic error of a percent ,
the correct phase transition temperature is obtained within the total error. Various such sources of systematic uncertainties are further explored in Appendix \ref{app:sys}.

\begin{table}
 \caption{\label{tab:4d}As Table \ref{tab:3d}, but for four dimensions.}
 \begin{tabular}{|c|c|c|c|c|c|}
 \hline
 Lattice    & $m_{\rm gribov}$ [GeV] & $a$ [GeV] & $b$ [GeV] & $\gamma_+$ & $\gamma_-$ \cr
 \hline
 $4\times46^3$ & 0.25$_{-2}^{-4}$ & 1.16$_{+-6}^{+3}$ & -0.04$_{-8}^{+200}$ & 1.26$_{+11}^{-16}$ & 0.19$_{+0.43}^{+378}$ \cr
 \hline
 $6\times48^3$ & 0.25$_{+2}^{-3}$ & 1.5$_{-3}^{+1}$ & -0.07$_{-17}^{+736}$ & 1.54$_{0.05}^{-12}$ & 0.6$_{+5}^{+45}$ \cr
 \hline
 \end{tabular}
\end{table}

\begin{figure}
\includegraphics[width=\linewidth]{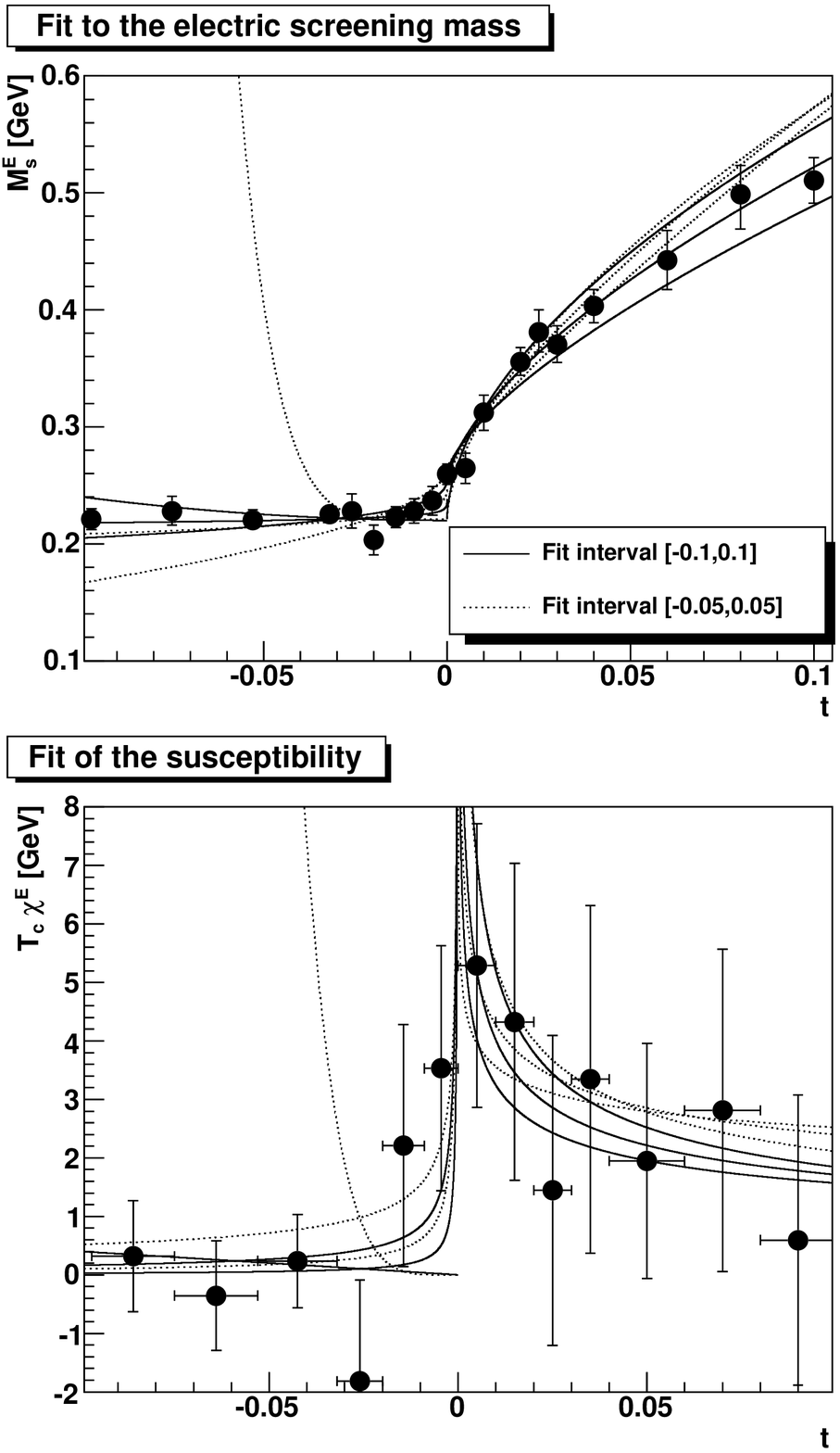}
\caption{\label{fig:4d-fit}The quality of the fits compared to the data from the $4\times46^3$ lattice. The standard fit uses the values from Table \ref{tab:3d}. As an alternative, a fit using only a more restricted reduced temperature interval $[-0.05,0.05]$ is also shown (dashed line), which yields $\gamma_-=0.5_{+6}^{+109}$ and $\gamma_+=1.5_{+2}^{-3}$. Fits have been performed on the screening mass using all data points, and the derivatives have then been calculated analytically.}
\end{figure}

It is possible to perform the same fits as for the three-dimensional
case with the Ansatz \pref{fit}. The results are given in Table
\ref{tab:4d}, and shown in Figure \ref{fig:4d-fit}. The situation is
qualitatively the same as in three dimensions. The resulting critical exponents
tend to be somewhat smaller than in three dimensions as expected from
universality. The temperature dependence of the electric screening
mass below $T_c$ is weaker than before, and in is fact within errors compatible with being esssentially constant. With critical scaling, the
prediction for its maximal slope at criticality would be roughly 
$\chi^E_\mathrm{max} \propto L^{0.6}$, which would predict a 25\%
difference between the two volumes used here. The somewhat stronger
volume dependence of the maximum as compared to the three dimensional
case might be reflected in Figure \ref{fig:4d-em} but this is well 
within the present errors still, and thus not significant.

A further point to consider in the four-dimensional case, in contrast
to the three-dimensional one, is renormalization. In three dimensions
the gluon propagator is finite, and no renormalization is
necessary. This is not the case in four dimensions. As a consequence,
the screening mass, in contrast to a pole mass, is
renormalization-scale dependent \cite{Maas:2011se}. Here, the
renormalization is performed by requiring $\mu^2D_L(\mu,T)=1$ at
$\mu=2$ GeV. However, this is problematic for two reasons. First, this
introduces a temperature dependence of the renormalization
constant. Secondly, the renormalization prescription for the gluon
propagator in Landau gauge is linked to the one for the
ghost propagator which might lead to a potential inconsistency, since
also the ghost propagator can acquire a temperature-dependent
renormalization. In the present case, however, these effects can be
estimated and are found to be well within the statistical errors, see
Appendix \ref{app:sys:renorm}. 

\subsection{Alternative observables}\label{sec:alt}

Since in both three and four dimensions the gluon propagator does not
show a behavior which can immediately be interpreted as a pure
critical behavior, a natural question might be whether one can
suppress the influence of non-thermal contributions depending on
whether they are due to discretization, finite volume artifacts,
genuine quantum effects, or the Gribov mass as mentioned. 

For this purpose the fluctuations of the screening mass can be used, which are defined as
\bea
m_s&=&\frac{1}{\sqrt{D^S(0)}}\label{sub}\\
D^S(0)&=&<D(0)>-\left<\sqrt{D(0)}\right>^2\nn.
\eea
\no This essentially amounts to determining the pure fluctuating part.

\begin{figure}
\includegraphics[width=\linewidth]{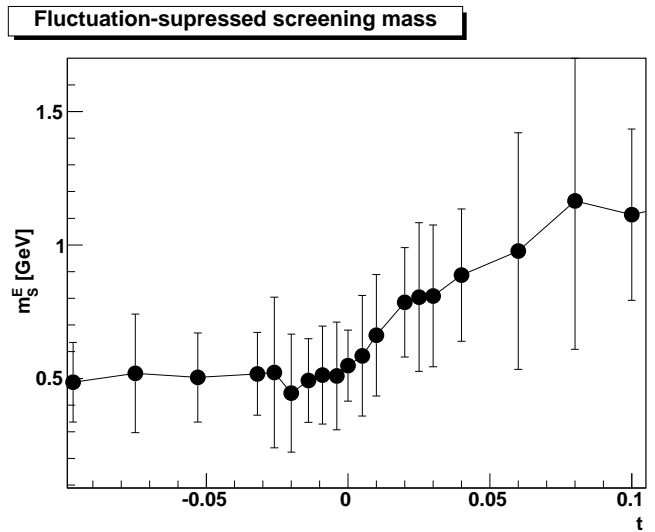}
\caption{\label{fig:sub}The fluctuation of the screening mass \pref{sub} in four dimensions from the $4\times46^3$ lattice.}
\end{figure}

The resulting screening mass for the four-dimensional case is shown in
Figure \ref{fig:sub}. There are a number of observations which can be
made, despite the significantly increased statistical
uncertainty. First, the global features of the screening mass are
maintained. Secondly, even after removing the non-fluctuating part
there is a non-zero baseline, which implies that the Gribov mass from
the gauge fixing is present even in the fluctuations, though, of
course its infinite volume behavior cannot be inferred from these
results. Thirdly, the slope again appears to increase at the phase
transition, which is not a statistically very sound result at present,
however. Nonetheless, the behavior is suggestive, and this would imply
a similar phase transition signal also in the fluctuations. 

In relation to the original goal of isolating critical behavior,
however, there is no sign of improvement as compared to the  
screening mass. 

A second alternative would be the magnetic screening mass. From the
results in \cite{Fischer:2010fx} and the results shown in Appendices
\ref{app:full} and \ref{app:stat}, it can already be inferred that it will not exhibit a
strong signal of the phase transition. Indeed, a more detailed
investigation of its susceptibility shows that, if there is any signal
of the phase transition encoded in the magnetic screening mass at all,
it is certainly substantially weaker than that in the electric
one. The same applies to the gauge-fixing sector in form of the
Faddeev-Popov ghosts. 

\section{A note on four-dimensional {\boldmath $SU(3)$} Yang-Mills theory}\label{sec:su3}

The physically more relevant case of four-dimensional $SU(3)$
Yang-Mills theory exhibits a (weak) first-order transition
\cite{Karsch:2001cy}. It has previously been studied with functional methods
\cite{Braun:2008pi,Braun:2010cy,Braun:2005uj,Braun:2007bx,Braun:2009gm,Chernodub:2007rn,Fischer:2009wc,Fischer:2009gk,Maas:2005hs,Maas:2004se,Gruter:2004bb,Maas:2011se,Roberts:2000aa,Zahed:1999tg,Zwanziger:2006sc,Marhauser:2008fz}
and on the lattice \cite{Maas:2011se,Aouane:2011fv,Fischer:2010fx}.
The correlation length remains finite, and
one does not expect critical scaling. In principle, one should be able
to study the finite-size scaling with integer exponents characteristic of 
first-order transitions which is rather expensive to investigate in lattice
simulations, however. 

\begin{figure}
\includegraphics[width=\linewidth]{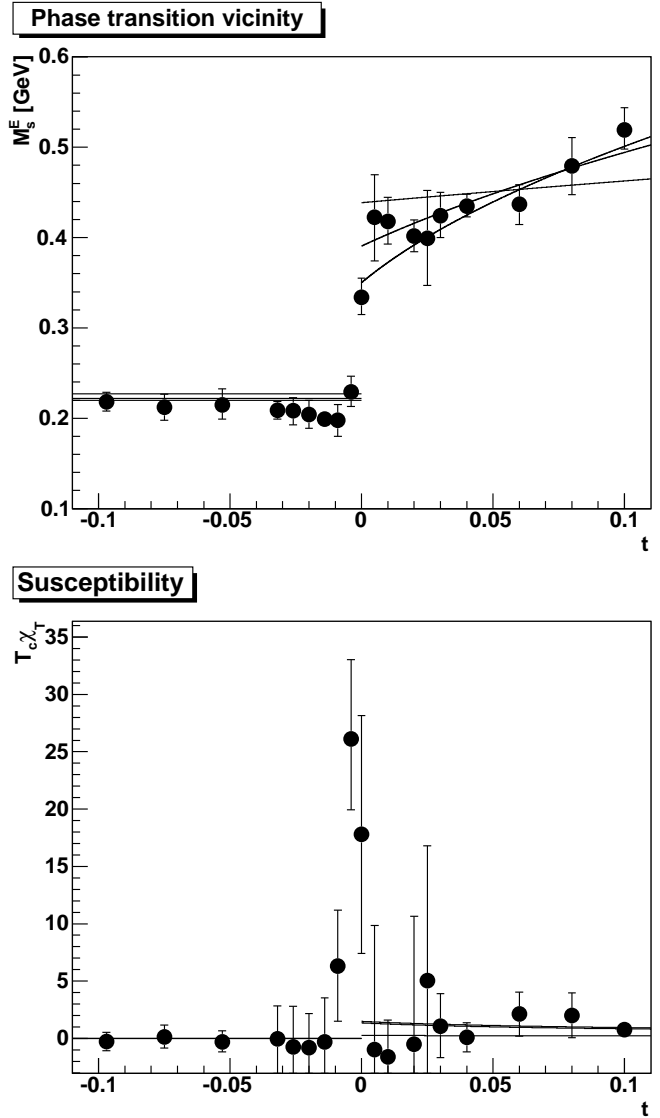}
\caption{\label{fig:su3}The electric screening mass and susceptibility for four-dimensional $SU(3)$ Yang-Mills theory on $4\times 34^3$ lattices close to the phase transition, compared to a fit of type \pref{eq:fo}.}
\end{figure}

Here we analyze the data from \cite{Fischer:2010fx,Maas:2011se}
supplemented by additionally generated temperature points and
with improved statistics in a closer window around the transition 
for comparison. This leads to an electric screening mass and a
susceptibility in four-dimensional $SU(3)$ as shown in Figure
\ref{fig:su3}. There is no significant temperature
dependence below $T_c$ in the range considered here. The constant
value serves as the baseline in our fit. Above $T_c $ we allow 
a temperature dependence of a form which is typical for a first-order
transition in addition to the constant contribution. With these
assumptions we observe that the data in the reduced temperature
interval $[-0.1,0.1]$ is well described by a fit of the form  
\be M_s^E=m_{\rm
  gribov}+\theta(t)\, a \sqrt{\delta+t }\label{eq:fo}, 
\ee
with parameters $m_{\rm gribov}=0.222_{+5}^{-3}$ GeV, $a=0.7_{+1}^{-4}$ GeV, and 
 $\delta=0.06^{-4}_{+37}$. 

The data provides rather clear indication that a
pronounced jump forms at the transition characteristic of its
first-order nature. A much more detailed analysis especially of the
finite volume behavior of these results will be necessary to
substantiate whether a discontinuity in the electric screening mass
develops in the infinite volume limit or not. The fact that the
transition is only weakly first order makes such an analysis
particularly demanding. At present we can conclude, however, that the
behavior looks significantly different from the second-order cases
discussed in Section \ref{sec:lat}.  An emerging discontinuity at the
phase transition appears to be indicative of a sensitivity of the
longitudinal gluon propagator to the order of the transition. One
should nevertheless be wary of lattice artifacts \cite{Aouane:2011fv}
in this exploratory study also. 

\section{Conclusions and outlook}\label{sec:sum}

In summary, we argue that critical behavior of the
confinement-deconfinement phase transition of pure Yang-Mills theory  
should be reflected in the longitudinal gluon
propagator, in particular, in the electric screening mass. Our data
shows in all cases, for $SU(2)$ and $SU(3)$, that the phase transition
can be clearly identified by a peak in its temperature derivative.  
Although it turns out to be rather difficult to disentangle a singular
critical contribution to the electric screening mass from a regular
one, which is presumably predominantly a Gribov copy effect, it seems
that there is such a contribution in $SU(2)$, and that this
contribution changes when going from three to four dimensions in a way
consistent with the change from the $2d$ to the $3d$ Ising
universality class. Our results for the critical exponents are still
somewhat off, however, which might be mainly due to an insufficient
way of isolating the corresponding contributions. Whether our
qualitative findings will finally be backed by a more quantitative
analysis on the basis of universality arguments, including a
verification of the expected finite-size scaling, as it is well
established for various other observables, will have to await further
study. 

In comparison, the electric screening mass of pure $SU(3)$ in four
dimensions appears to show the discontinuity indicative of the 
first-order phase transition there. This seems to indicate that the
electric gluon propagator is sensitive to the order of the transition,
likewise. 

To ultimately clarify all this will require both substantially larger
spatial volumes as well as finer discretizations. It will be important to 
further pursue this for two reasons: One is that identifying critical
behavior in this simplest of observables will improve our
understanding of the driving mechanisms behind critical behavior and
the order of the phase transition in general. Secondly, this clarification is of
great interest in the construction of realistic truncation schemes
for functional methods which have the potential of providing    
a genuine first-principle approach to describe full QCD with various
quark flavors and masses at finite temperature and density
\cite{Pawlowski:2010ht}. 

Thus, to further pursue these investigations is of prime importance in 
our efforts to combine both numerical lattice and functional continuum
methods for a reliable determination of the QCD phase diagram.\\

\no{\bf Acknowledgments}\\

We are grateful to Christian S.\ Fischer and Michael M\"uller-Preussker for a critical reading of the manuscript.

This work is supported by the Helmholtz Alliance HA216/EMMI and the
Helmholtz International Center for FAIR within the LOEWE program of
the State of Hesse.  A.M. was supported by the FWF under grant
number M1099-N16 and by the DFG under grant number MA
3935/5-1. L.v.S. received additional support from the Helmholtz
Association, Grant VH-NG-332, and the European Commission,
FP7-PEOPLE-2009-RG, No.~249203. D.S. acknowledges 
support by the Landesgraduiertenf\"orderung Baden-W\"urttemberg via
the Research Training Group ``Simulational Methods in Physics''.  The
numerical simulations were carried out on bwGRiD
(http://www.bw-grid.de), member of the German D-Grid initiative, on
the computer cluster of the ITP, University of Heidelberg, and the HPC
clusters of the Universities of Graz and Jena. The ROOT framework
\cite{Brun:1997pa} has been used in this project.

\appendix

\section{Full momentum dependence}\label{app:full}

Besides the critical behavior manifest at zero momentum also the full momentum dependence of the propagators is important, both for combining with functional methods as well as to estimate the reliability of other approximation techniques, like e.\ g.\ hard-thermal-loop calculations close to the phase transition.

Hence, in this appendix the full momentum dependence of both parts of the propagator, the electric and the magnetic one, for the zeroth Matsubara frequency will be presented. Higher Matsubara frequencies can then be approximated rather accurately with this information \cite{Fischer:2010fx}. Results for the Faddeev-Popov ghost can be found elsewhere \cite{Maas:2011se,Cucchieri:2007ta,Fischer:2010fx,Maas:unpublished}.

\begin{figure}
\includegraphics[width=\linewidth]{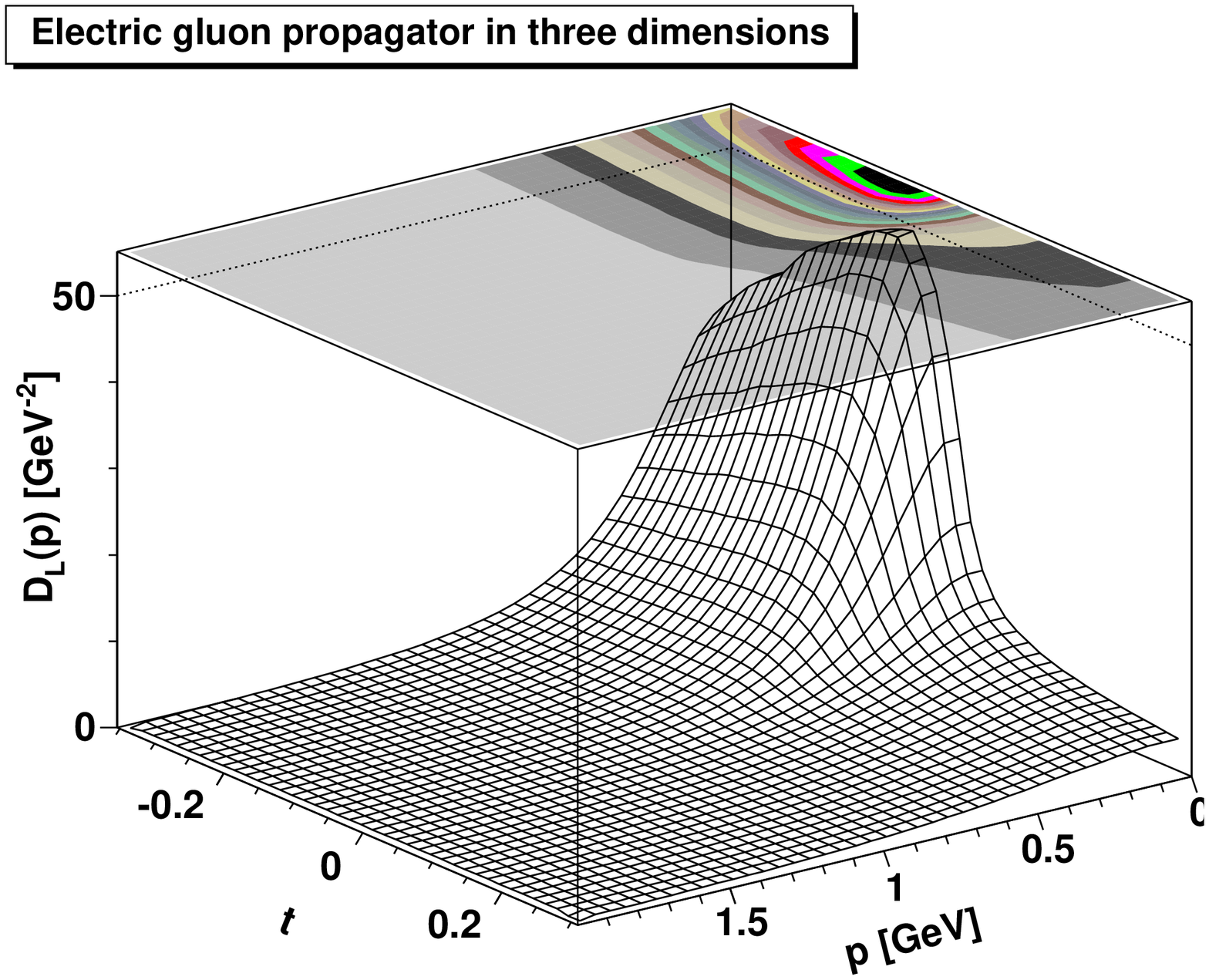}\\
\includegraphics[width=\linewidth]{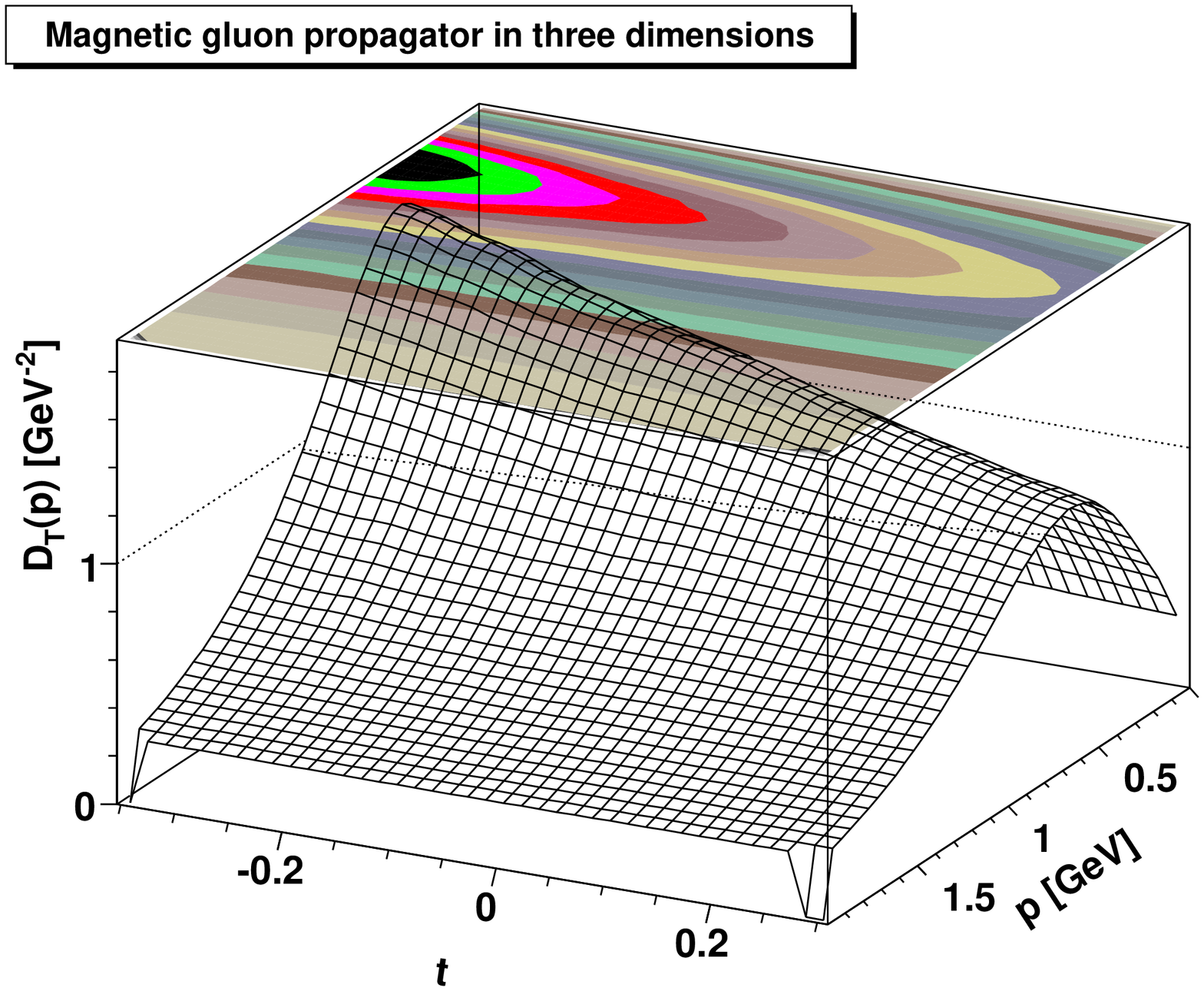}
\caption{\label{fig:3d-gp}The full momentum dependence of the zero-mode electric (top panel) and magnetic (bottom panel) gluon propagator in three dimensions on a $4\times 128^2$ lattice.}
\end{figure}

\begin{figure}
\includegraphics[width=\linewidth]{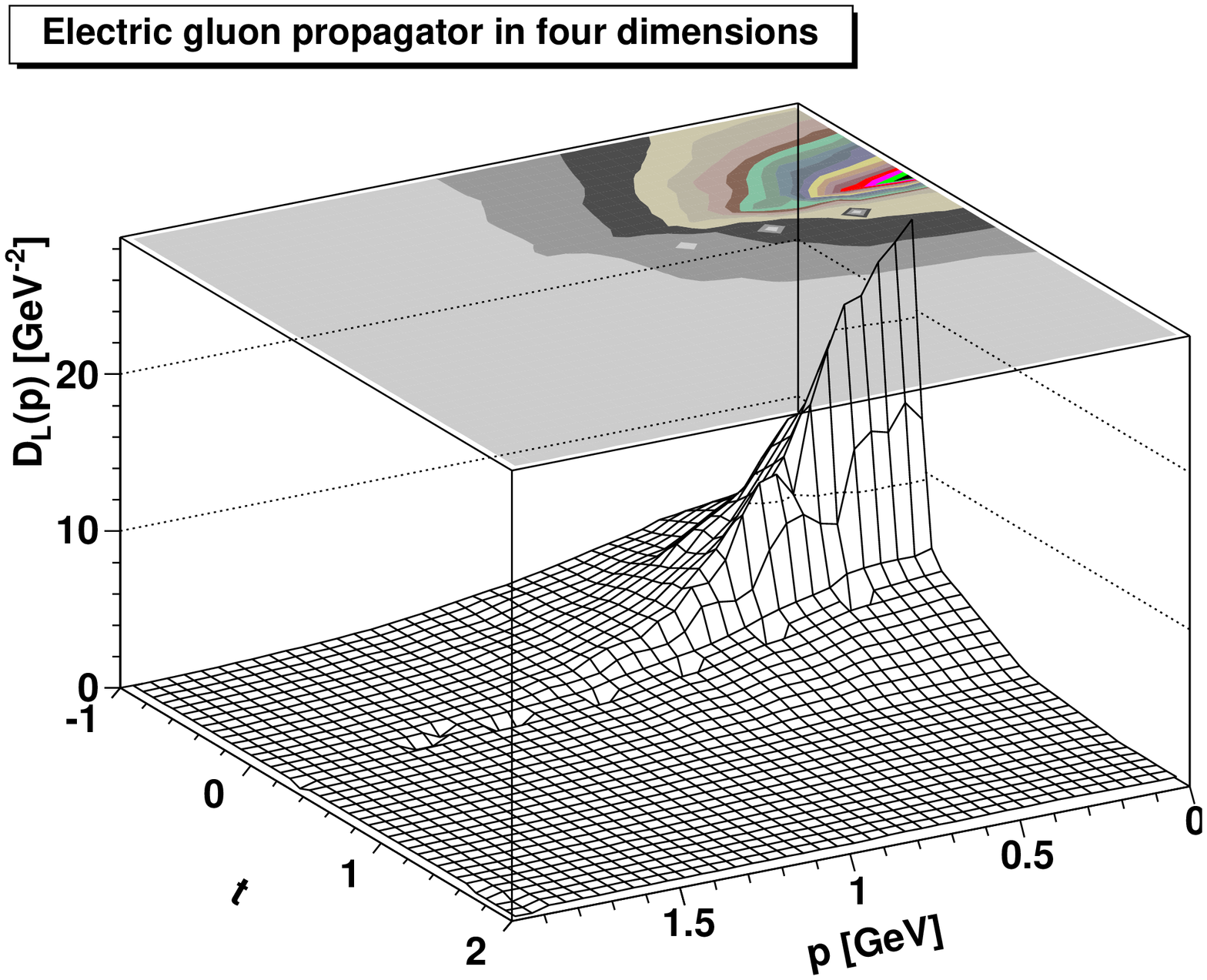}\\
\includegraphics[width=\linewidth]{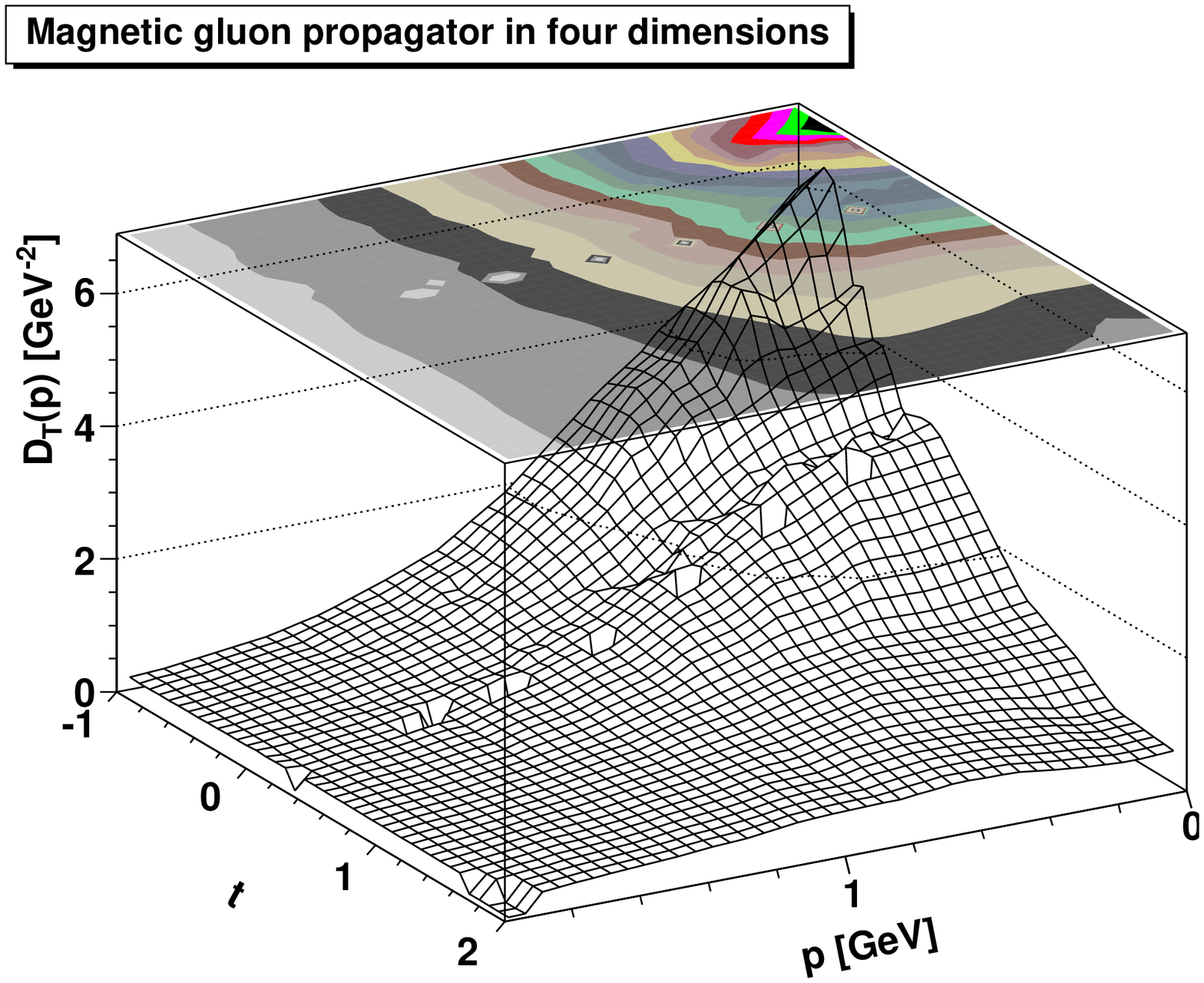}
\caption{\label{fig:4d-gp}The full momentum dependence of the zero-mode electric (top panel) and magnetic (bottom panel) gluon propagator in four dimensions. To extend the temperature range outside the $[-0.1,0.1]$ interval additional lattice sizes have been used, which are chosen to have an approximately constant number of lattice points as the $6\times 48^3$ lattices in this reduced temperature range. Dips along the $t=0$ line are artifacts from the graphical interpolation routine.}
\end{figure}

The results in three dimensions are shown in figure \ref{fig:3d-gp} for the $4\times128^2$ lattice. For four dimensions the results for the $6\times48^3$ lattice are shown in figure \ref{fig:4d-gp}. The results for the $4\times46^3$ lattice can be found in \cite{Fischer:2010fx}. For both dimensions, the results are qualitatively similar, up to the differences present already at zero temperature. In particular, the abrupt change in the electric screening mass discussed in the main text finds its echo at finite momentum. The magnetic propagator shows no pronounced dependence on the temperature, except for a general suppression.

It is an interesting observation that the smoothness of the electric screening mass in three dimensions compared to four dimensions also surfaces in the finite momentum behavior. This again can be interpreted as a sign that Yang-Mills theory in four dimensions is 'closer' to a first order transition than three dimensions, as is known from other observables.

\section{Systematic errors}\label{app:sys}

Calculations like the present ones are sensitive to various systematic errors, as has been pointed out repeatedly
\cite{Cucchieri:2007ta,Fischer:2010fx,Bornyakov:2010nc,Cucchieri:2011di,Aouane:2011fv}. In the following various such errors will be discussed in turn, including the scale setting, renormalization, finite-volume effects and aspect ratio issues.

\subsection{Scale setting}\label{app:sys:scale}

The scale in the present calculations has been set in four dimensions using an interpolation formula of the type, 
\be
a\sqrt{\sigma}=\frac{c_1}{\beta}+\frac{c_2}{\beta^2}+\frac{c_3}{\beta^3}+\frac{c_4}{\beta^4}\nn.
\ee
with $\sigma$, the string tension, set to $(440 $MeV$)^2$, and where the coefficients have been fitted to the results from \cite{Fingberg:1992ju}, yielding $c_1=31.9$, $c_2=-237$, $c_3=574$, and $c_4=-444$. However, since reduced temperatures of size 0.01 have been included in the calculation, this setting maybe too imprecise. Using data from \cite{Lucini:2003zr}, different values for the coefficients have been obtained as $c_1=33.6$, $c_2=-249$, $c_3=599$, and $c_4=-461$. In this case, the critical $\beta$ was determined more precisely as $\beta=2.2986$, instead of $\beta=2.299$, as used in the main part of the text. However, this leads only to a shift in $t$ of the order of 0.01, i.\ e., reduced temperatures in the main text should be taken to have a systematic uncertainty of $\pm 0.01$. This accommodates the shift of the peak of the susceptibilities in figure \ref{fig:4d-em} compared to $t=0$. Similar comments apply, of course, to the three-dimensional case.

\subsection{Renormalization}\label{app:sys:renorm}

In four dimensions, the propagators have to be renormalized. As a consequence, also the screening mass is renormalized by the same wave-function renormalization factor. In the main text, this is done at each temperature individually at 2 GeV. Of course, since finite temperature is not introducing any new divergences \cite{Das:1997gg}, it is also possible to renormalize once and for all at zero temperature. However, since the renormalization constants are both potentially volume-dependent and not available for all $\beta$ values investigated, this has not been done here.

\begin{figure}
\includegraphics[width=\linewidth]{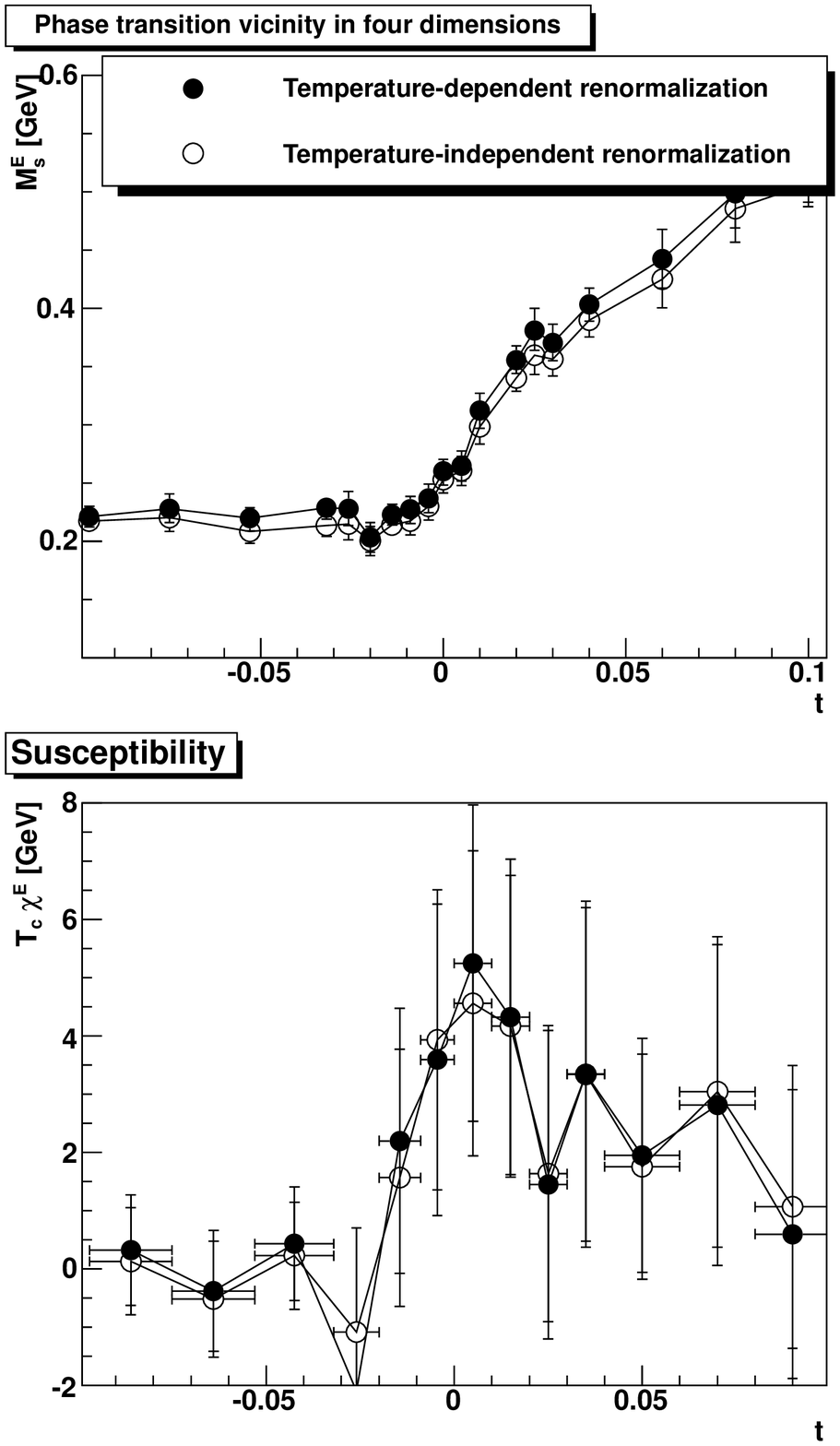}
\caption{\label{fig:renorm}The electric screening mass and susceptibility determined using a temperature-dependent renormalization prescription, as in the main text, and a temperature-independent renormalization prescription.}
\end{figure}

To assess the relevance of renormalization, it is shown in figure \ref{fig:renorm} the effect of renormalizing only once at zero temperature for the $4\times 46^3$ lattice. Here, the $\beta$ dependence of the renormalization constants is ignored, given that it is expected to be only logarithmic in the scaling regime. The result show indeed no change within statistical errors. This was expected, as at a momentum of 2 GeV temperature effects are still rather small. This may change at lager temperatures, but is not relevant for the present purpose.

\subsection{Finite volume effects}

\begin{figure}
\includegraphics[width=\linewidth]{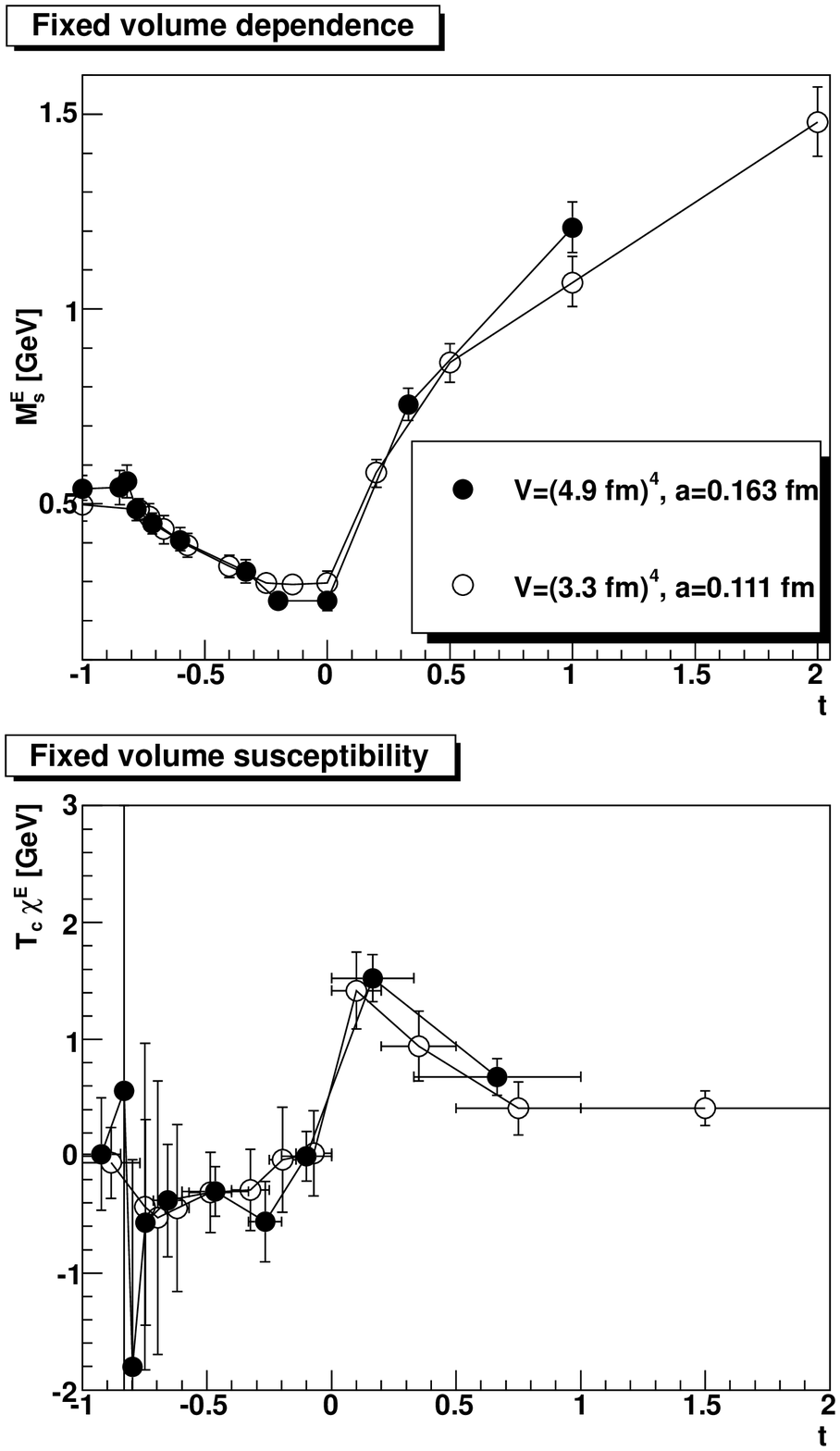}
\caption{\label{fig:sys-v}The electric screening mass and susceptibility as a function of the reduced temperature at fixed spatial volume for two different volumes.}
\end{figure}

In the main text, the spatial volume was not kept fixed, but small variations have been permitted to obtain a very dense mesh of temperatures around the critical one. In particular, the spatial volumes at very low temperatures and very high temperatures are quite different. To assess the influence of this effect, the electric screening mass and susceptibility it shown at fixed spatial volume in figure \ref{fig:sys-v}.

As is visible, the overall generic behavior over the whole temperature range is not significantly altered when using a fixed instead of a varying spatial volume. However, as has already been observed in other calculations \cite{Cucchieri:2011di}, it is visible that a finer lattice seems to increase the plateau directly before the phase transition. Of course, in the present case neither the temperature mesh, nor the spatial volumes, are exceptionally large. Therefore, this should be taken only as an indication, and requires further study.

\subsection{Aspect ratio}

\begin{figure}
\includegraphics[width=\linewidth]{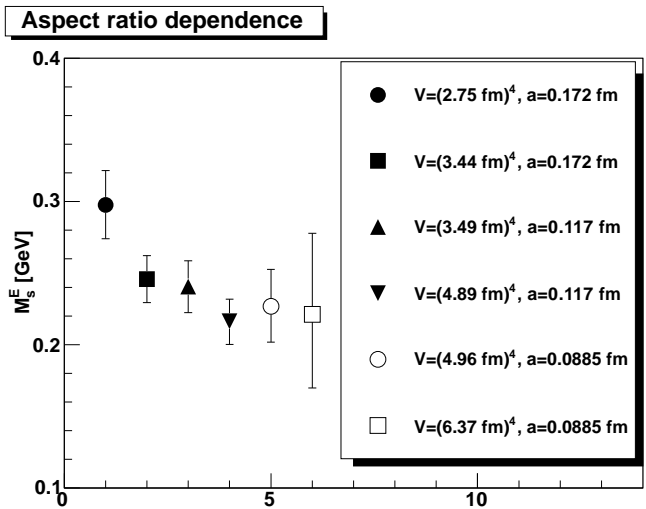}
\caption{\label{fig:sys-ar}The electric screening mass as a function of the aspect ratio at $T=0.95T_{c}$. The aspect ratios from left to right are, respectively, 1:4, 1:5, 1:5, 1:7, 1:7, and 1:9.}
\end{figure}

As already indicative in the discussion of the finite volume effects, also the aspect ratio and the temporal discretization play a significant role. In figure \ref{fig:sys-ar} the electric screening mass at a fixed temperature is shown for a number of different spatial volumes and aspect ratios. In the figure, the aspect ratio increases from left to right, approaching the infinite-volume and continuum limit such that the temperature extension remains finite, while the spatial extension diverges. It is visible that at small aspect ratios and volumes a significant dependence on these two parameters exist. However, towards the infinite-volume and continuum limit, this dependency quickly diminishes, if the limit is taken in the presented order, though the statistical error bars are rather large. The final value is in agreement with the one obtained in the main text. In contrast, at fixed aspect ratio and spatial volume a much larger impact of going to the continuum limit has been found \cite{Cucchieri:2007ta,Cucchieri:2011di}, indicating that the different approaches towards the desired limit have to be investigated carefully.

\section{Statistical errors}\label{app:stat}

\begin{figure}[ht]
\includegraphics[width=\linewidth]{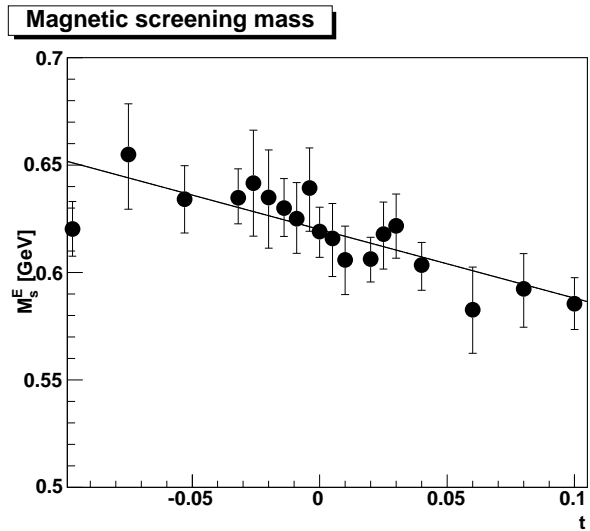}
\caption{\label{fig:mm}The magnetic screening mass on the $4\times 46^3$ lattice close to the phase transition temperature compared to a linear fit.}
\end{figure}

The assessment of the quality of fits presented in the main text is a rather complicated problem. Usual $\chi^2$ checks can only be expected to give a reliable estimate if the fit form is linear, but the present one is non-linear, non-analytic, and non-continuously differentiable, and in the case of SU(3) not even continuous. To estimate the statistical error of the fit thus instead a comparison to a null hypothesis is used. This null hypothesis is the absence of any signal of the phase transition, i.\ e., a trivial linear behavior
\be
M_s^E(t)=m_0+a(1+t)\label{mmfit}.
\ee
\no For a quantity not influenced strongly by the phase transition this is an adequate description, as is shown in case of the magnetic screening mass in figure \ref{fig:mm}.

To estimate the error, the results for the electric screening mass have also been fitted to the formula \pref{mmfit}. Then, instead of using the average values for the electric screening mass, a value within the 67\% confidence interval for each temperature point was chosen such that it was closest to the fitted linear behavior. Then these points were fitted again within the fit ansatz \pref{eq:fullmass}. Since the ansatz \pref{eq:fullmass} contains also this null hypothesis, this gives exponents as close as possible to the null hypothesis. To estimate an opposite error, the fluctuation were performed towards as far away as possible from the null hypothesis \pref{mmfit}. These two fits gave two further fit parameters. In the tables \ref{tab:3d} and \ref{tab:4d} the deviation of the fit parameters towards the null hypothesis is set as a subscript, while those as far away from the null hypothesis as possible are set as superscripts. These results have been used in the figures \ref{fig:3d-fit} and \ref{fig:4d-fit}. In the four-dimensional case in the low-temperature phase the uncertainties in the fits is quite significant, a consequence of the compatibility with a constant within errors. Thus fits with free pre-factor and exponent can turn out to be optimal by combining very extreme combinations of both. The result should thus be rather interpreted as that the low-temperature behavior is within errors compatible with a constant.

In the SU(3) case in figure \ref{fig:su3} the error for the fit was determined in the same way.

\bibliographystyle{bibstyle}
\bibliography{bib}


\end{document}